\documentclass[aps,prl,twocolumn,superscriptaddress,showpacs,amssymb,amsmath]{revtex4-1}

\usepackage{amssymb}

\usepackage{graphics}
\usepackage{graphicx}
\usepackage{epsfig}
\usepackage{amssymb}

\begin{document}


 \title{Influence of the spin-orbit interaction in the impurity-band
states of n-doped semiconductors}

\author{Guido A.\ Intronati}
\author{Pablo I.\ Tamborenea}
\affiliation{Departamento de F\'{\i}sica and IFIBA, FCEN, 
Universidad de Buenos Aires, Ciudad Universitaria, Pab.\ I, 
C1428EHA Ciudad de Buenos Aires, Argentina}
\affiliation{Institut de Physique et Chimie des Mat\'{e}riaux de Strasbourg, 
UMR 7504, CNRS-UdS, 23 rue du Loess, BP 43, 67034 Strasbourg Cedex 2, France}

\author{Dietmar Weinmann}
\author{Rodolfo A.\ Jalabert}
\affiliation{Institut de Physique et Chimie des Mat\'{e}riaux de Strasbourg, 
UMR 7504, CNRS-UdS, 23 rue du Loess, BP 43, 67034 Strasbourg Cedex 2, France}






\begin{abstract}
We study numerically the effects of the Rashba spin-orbit interaction
on the model of electrons in n-doped semiconductors of Matsubara and
Toyozawa (MT).
We focus on the analysis of the density of states (DOS) and the 
inverse participation ratio (IPR) of the spin-orbit perturbed states 
in the MT set of energy eigenstates in order to characterize the eigenstates
with respect to their extended or localized nature.
The finite sizes that we are able to consider necessitate an enhancement of 
the spin-orbit coupling strength in order to obtain a meaningful perturbation.
The IPR and DOS are then studied as a function of the enhancement
parameter.
\end{abstract}

\maketitle

The metal-insulator transition (MIT) is one of the paradigms of Condensed 
Matter Physics \cite{mot,mottnewer} and new features constantly appear 
according to the physical properties under study, the specific system or the
emerging experimental techniques. 
The richness of the physics around the MIT stems from the fact that it is 
a quantum phase transition where disorder and Coulomb interactions coexist 
and compete in the determination of the ground state. 
In the case of the n-doped semiconductors, the MIT appears at doping densities 
where the Fermi level is in the impurity band \cite{shk-efr}. 
This observation allows a description taking into account only the electronic
states built from the hydrogenic ground state of the doping impurities. 
For densities slightly larger than the critical one (in the metallic side of the
transition) non-interacting models, like the Matsubara-Toyozawa (MT) 
\cite{mat-toy} are applicable. Furthermore, the previous description in terms
of impurity sites is often traded by the Anderson model of a tight-binding
lattice with on-site or hopping disorder. 
Large amounts of numerical work have been devoted to the Anderson model 
\cite{kra-mac} and the critical exponents obtained fit reasonably well 
those of the experimental measurements \cite{kra-mac-oht-sle}.

The recently developed field of spintronics is contributing to put the 
MIT again into the focus of the condensed matter community. 
A key concept for possible applications of spintronics is the spin 
relaxation time, that is, the typical time in which the electron spin 
loses its initially prepared direction. 
Interestingly, the maximum spin relaxation times in n-doped semiconductors 
have been observed for impurity densities close to that of  the MIT 
\cite{zar-cas,kik-aws,dzh,sch-hei-roh,roe-ber-mue}.
This intriguing physics is not completely understood at present, and various 
mechanisms of spin relaxation have been thought to be active at the MIT 
region \cite{shk,kav,put-joy,ISA1}.
At the level of models, the generalization of the Anderson model in order to 
include some spin-orbit coupling has been provided by Ando \cite{and}.
While this model is very useful to study the progressive breaking of the spin
symmetry \cite{asa-sle-oht}, its connection with experimentally relevant
systems requires the estimation of coupling parameters which are not
obtainable from first principles.
This situation has lead us to reconsider the MT model of impurity sites
randomly placed in order to incorporate in it the spin-orbit interaction.
The various spin-orbit couplings (intrinsic and extrinsic) can be included 
and lead to effective Hamiltonians which depend on fundamental material
constants, rather than on adjustable parameters.

In this paper we first consider the MT model in order to characterize the
regions of extended and localized states, analyzing the limitations of the
model and the conditions of applicability. 
We then include one of the sources of spin-orbit coupling, the Rashba
interaction \cite{ISA1} to study how the previous picture evolves 
under increasing values of the coupling strength. 
This work is a necessary step towards the understanding of spin dynamics 
in the generalized models that will allow us to extract the spin 
relaxation times close to the MIT. 

We start by considering the MT Hamiltonian \cite{mat-toy}
\begin{equation}\label{eq:MT}
\mathcal{H}_0 = \sum_{m\neq m',\sigma} t_{m' m}^{\sigma \sigma}
                                    c_{m' \sigma}^{\dag}  c_{m \sigma},
\end{equation}
where $c_{m' \sigma}^{\dag}$ ($c_{m \sigma}$) represents the creation
(annihilation) operator for the ground state of the impurity at
site $m'$ ($m$) with spin projection $\sigma$ in the $z$-direction. 
The spin degree of freedom is irrelevant for the MT model, but it will become
crucial later. 
The hopping matrix element is
\begin{eqnarray}\label{eq:t_MT}
t_{m' m}^{\sigma \sigma} &=&
\langle \phi_{m'}|V_{m'}|\phi_{m}\rangle \nonumber \\
&=& -V_0 \left(1+\frac{R_{m'm}}{a} \right) 
\exp{\left(-\frac{R_{m'm}}{a}\right)} \, ,
\end{eqnarray}
where $\phi_p({\mathbf{r}})=\phi(|\mathbf{r}-\mathbf{R}_p|)$, with
$\phi(\mathbf{r})=1/(\pi a^3)^{1/2} \times$ \\ $\exp{(-r/a)}$, 
and $a$ is the effective Bohr radius. 
The Coulombic potential produced by the impurity placed at 
$\mathbf{R}_p$ is $V_p(\mathbf{r})=-V_0(a/|\mathbf{r}-\mathbf{R}_p|)$, 
where $V_0=e^2 /\varepsilon a$ and $\varepsilon$ stands for the static
dielectric constant of the semiconductor. 

In order to characterize the electronic eigenstates in the impurity band 
from the point of view of their extended or localized nature, we solve
numerically the eigenvalue problem defined by the Hamiltonian (\ref{eq:MT})
for given configurations in which $N$ impurities are randomly placed in a 
three-dimensional volume. 
Performing the impurity average we obtain (Fig.\ 1) the density of 
states (DOS) and the inverse participation ratio (IPR) for three densities 
on the metallic side of the transition. 
The impurity band develops around the $E=0$ level of the isolated impurity 
in an asymmetric fashion: the DOS exhibits a long low-energy tail while 
the high-energy part is bounded by $E=1$ (in units of $V_0$). 
We verify that the width of the impurity band increases with the doping density.
The highest-energy states correspond to electronic wave functions localized on
small clusters of impurities. 
This clustering is known to happen in realistic systems due to the lack of 
hard-core repulsion between impurities on the scale of $a$ \cite{tho-etal,ISA1}.

\begin{figure}
\includegraphics[width=\linewidth]{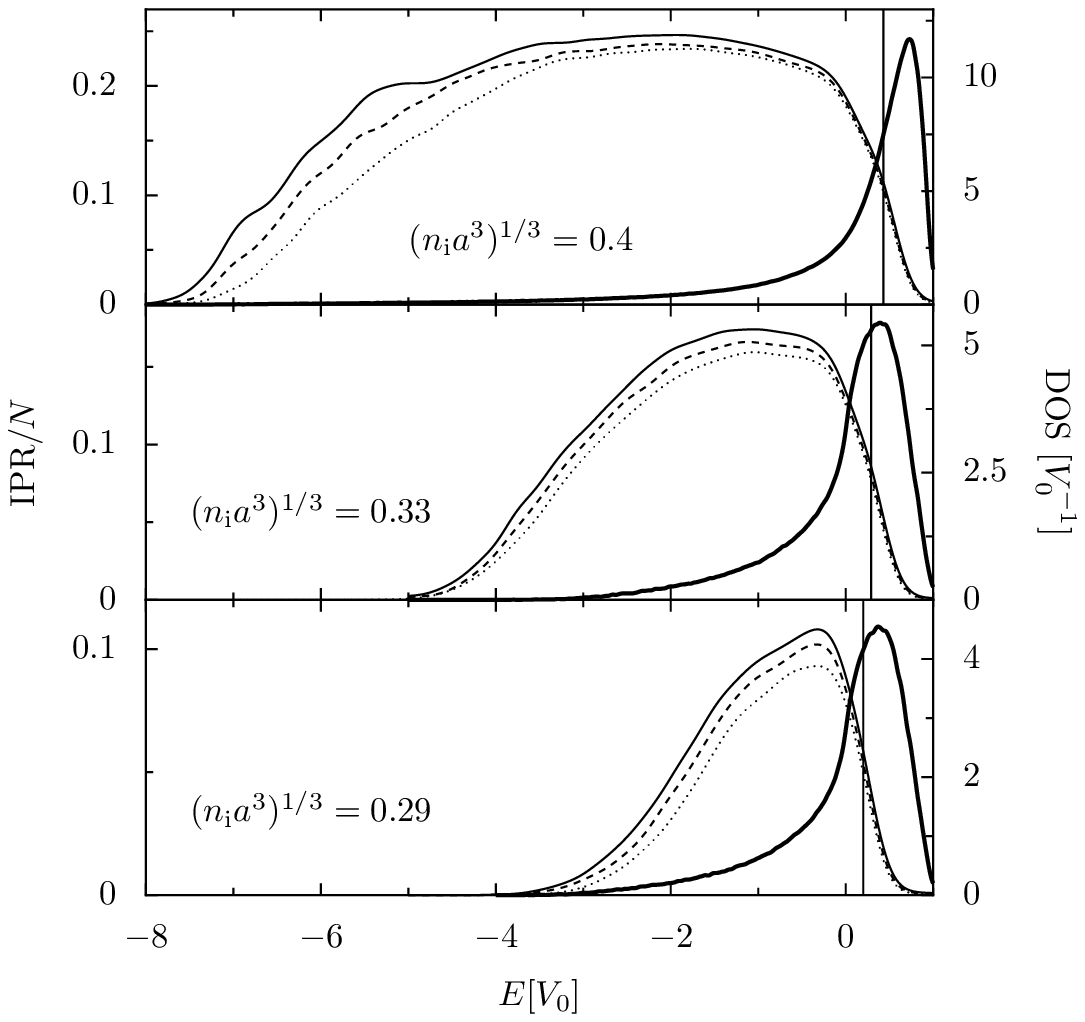}
\caption{Density of states (DOS, thick line and right scale) and inverse 
participation ratio (IPR, left scale) for three different densities on the 
metallic side of the metal-insulator transition, obtained through impurity 
averaging in the Matsubara-Toyozawa model.
The solid, dashed and dotted curves of $\mbox{IPR}/N$ are for $N=2744, 
4096$ and $5832$, respectively, and the vertical lines indicate the Fermi 
energy.}
\label{fig:IPRDOSMT}
\end{figure}

Before continuing with the analysis of the numerical results obtained from 
the MT model, we discuss some technical features of the model and the 
difficulties that we face in trying to improve upon it. 
Firstly, we notice that the chosen basis set is not orthogonal. 
In principle, we can deal with this issue by writing a generalized eigenvalue 
problem which includes the matrix of orbital overlaps \cite{maj-and,chi-hub}. 
This procedure results in unphysical high-energy states (with $E\gg 1$) that 
necessitate the inclusion of hydrogenic states beyond the $1s$ orbital in order 
to be described properly.
However, care must be taken since enlarging the basis set leads to the problem 
of overcompleteness.
Fortunately, for the properties we are interested in, the effects arising from 
non-orthogonality are known to be small for moderate doping densities, and 
that is why we will not consider them in the numerical work, thus staying
within the original MT model. 
Finally, another drawback  of the MT model is that the high-energy edge of 
the impurity band overlaps with the conductance band, which starts at $V_0/2$ 
(the effective Rydberg) and is not included in the MT description. 
As seen in Fig.\ \ref{fig:IPRDOSMT} the DOS beyond $V_0/2$ is 
always very small, and therefore we can ignore the effects that the 
hybridization of the bands would yield in a more complete model.  

The IPR measures the degree of extension of the electronic wave function.
Large values of the IPR denote in general extended states while low values
are associated with localized states.
In particular, an homogeneously distributed wave function would have 
an $\mbox{IPR}/N=1$, while a localized state would exhibit a vanishing 
$\mbox{IPR}/N$ upon increasing values of $N$.
In our study of spin relaxation it is important to know whether the states
near the Fermi energy are in the region of extended or localized states.

The determination of the mobility edges from the size scaling of 
Fig.\ \ref{fig:IPRDOSMT} is not straightforward.
This difficulty arises from the heavily structured DOS of the MT model.
At low energy the small values of the DOS translates into a poor
statistics for feasible sizes.
In the high-energy part of the impurity band the separation between the
curves corresponding to different values of $N$ is masked by the small
values of the $\mbox{IPR}/N$.

For the highest density (top panel) the $\mbox{IPR}/N$ exhibits a relatively
flat region at intermediate energies, which is approximately independent
of $N$ for the two largest system sizes.
The lower mobility edge can be located roughly at $E \sim 3.5$, where the
latter curves separate.
For lower impurity densities (lower panels) the previous analysis becomes
increasingly demanding in terms of system sizes.
We see that the flat region of $\mbox{IPR}/N$ shrinks from which we can 
conclude that the lower mobility edge is shifting towards higher values of $E$.

The study of spin relaxation in doped semiconductors with densities close 
to the that of the MIT calls for a generalization of the previously discussed
MT model that incorporate spin-orbit coupling. 
Such an extension was done in Ref.\ \cite{ISA1}, where a spin-flip term
\begin{equation}\label{eq:ISA}
{\mathcal H}_1 = \sum_{m\neq m',\sigma} t_{m' m}^{\sigma \overline{\sigma}} \
c_{m' \overline{\sigma}}^{\dag} \ c_{m \sigma}
\end{equation}
was added to $\mathcal{H}_0$ ($\overline{\sigma} = -\sigma $).
Similarly to the spin-conserving case, we have
\begin{equation}\label{eq:t_spin_flip}
t_{m' m}^{\sigma \overline{\sigma}} = \sum_{p \neq m}
\langle \tilde{\psi}_{m' \overline{\sigma}}|V_p|\tilde{\psi}_{m \sigma}
\rangle.
\end{equation}
The wave function $\tilde{\psi}_{m \sigma}$ is a spin-mixed conduction-band
state with an envelope part $\phi_{m}(\mathbf{r})$ and a lattice-periodic
part which has a small spin admixture.
In Ref.\ \cite{ISA1}, the expression of the matrix elements of 
Eq.\ (\ref{eq:t_spin_flip}) within an 8-band Kane model were obtained.
Since the two-center integrals $p=m, m'$ were shown to vanish, the spin-flip
hopping amplitude is given by the sum of three-center integrals with 
$p \neq m, m'$.
The three-center integrals cannot be analytically solved in general.
In Ref.\ \cite{ISA1} we provided approximate analytical expressions of 
$t_{m' m}^{\sigma \overline{\sigma}}$ which allowed us to estimate the spin relaxation times.
On the other hand, in this work we take the route of the numerical
evaluation of the three-center integrals.

Spin-orbit coupling is known to favor the delocalization of disordered
systems in two dimensions.
In what follows we repeat the previous analysis, done for the MT model, 
for the spin-orbit generalized model in this case, for increasing values 
of the spin-orbit coupling $R_r$ (the subindex $r$ stands for Rashba). 

\begin{figure}
\includegraphics[width=6cm]{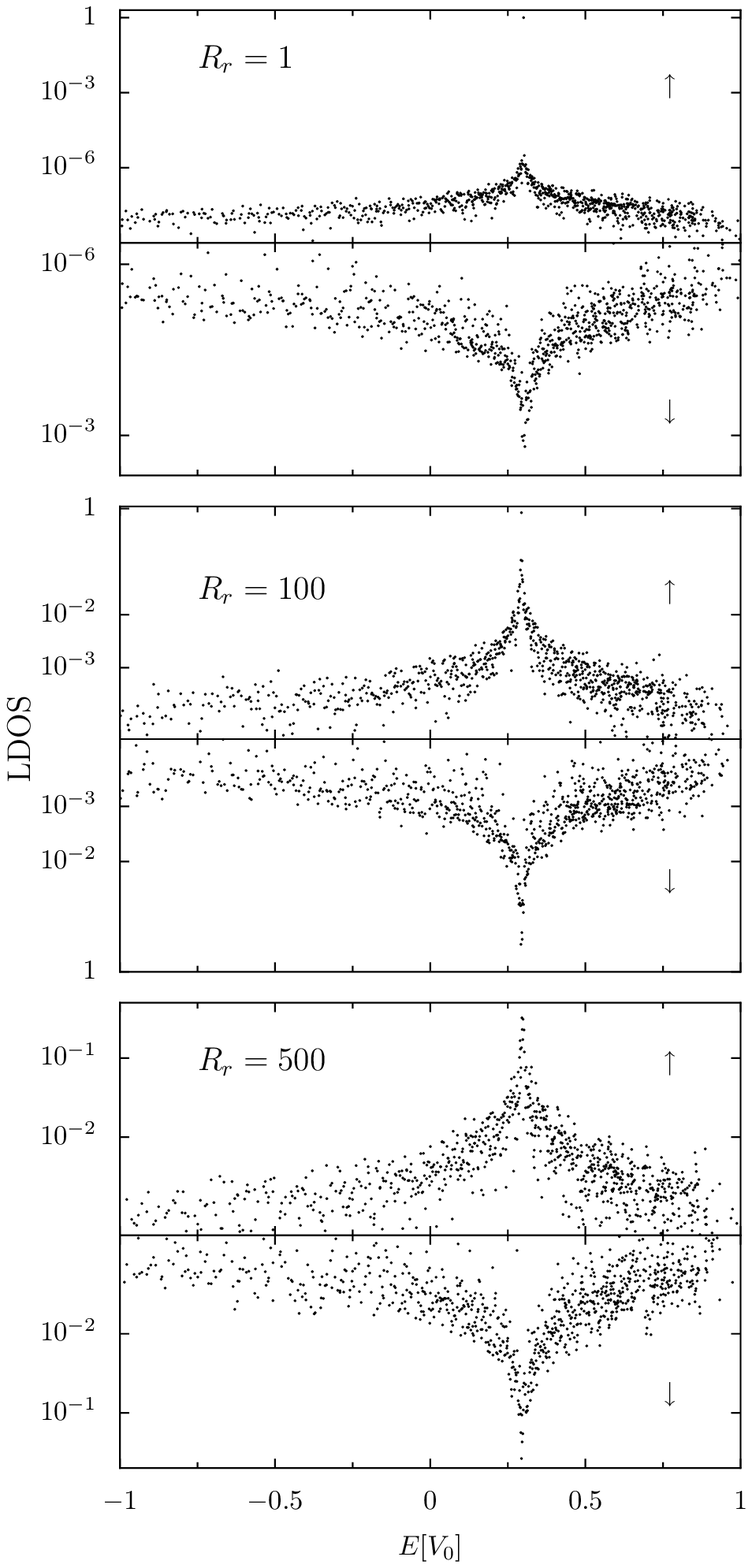}
\caption{Spectral decomposition of a Matsubara-Toyozawa eigenstate
into the basis set formed by the eigenstates of our spin-orbit extended
model.
The system size is $N=1000$ and the density is given by 
$(n_i a^3)^{1/3} = 0.33$.
Top, middle, and bottom panels correspond to $R_r=1, 100, 500$, respectively.
}
\label{fig:spectraldecomp}
\end{figure}

For the realistic values of the spin-orbit coupling strength ($R_r=1$) 
considered in Ref.\ \cite{ISA1}, the spin-admixture perturbation energies
are, even for largest system sizes that we can treat numerically, 
orders of magnitude smaller than the MT level spacing.
In Fig.\ \ref{fig:spectraldecomp} we show the spectral decomposition 
of a MT eigenstate with $\sigma=1$ in the basis of spin-admixed eigenstates 
of $\mathcal{H}_0+\mathcal{H}_1$.
Only for enhanced values of $R_r$ do we obtain significant projections 
into the two spin-admixed subspaces.

The study of larger values of $R_r$ then appears not only as a useful
tool for analyzing the progressive inclusion of spin-orbit effects, but 
also as a need for numerical simulations of the spin dynamics.

\begin{figure}
\includegraphics[width=\linewidth]{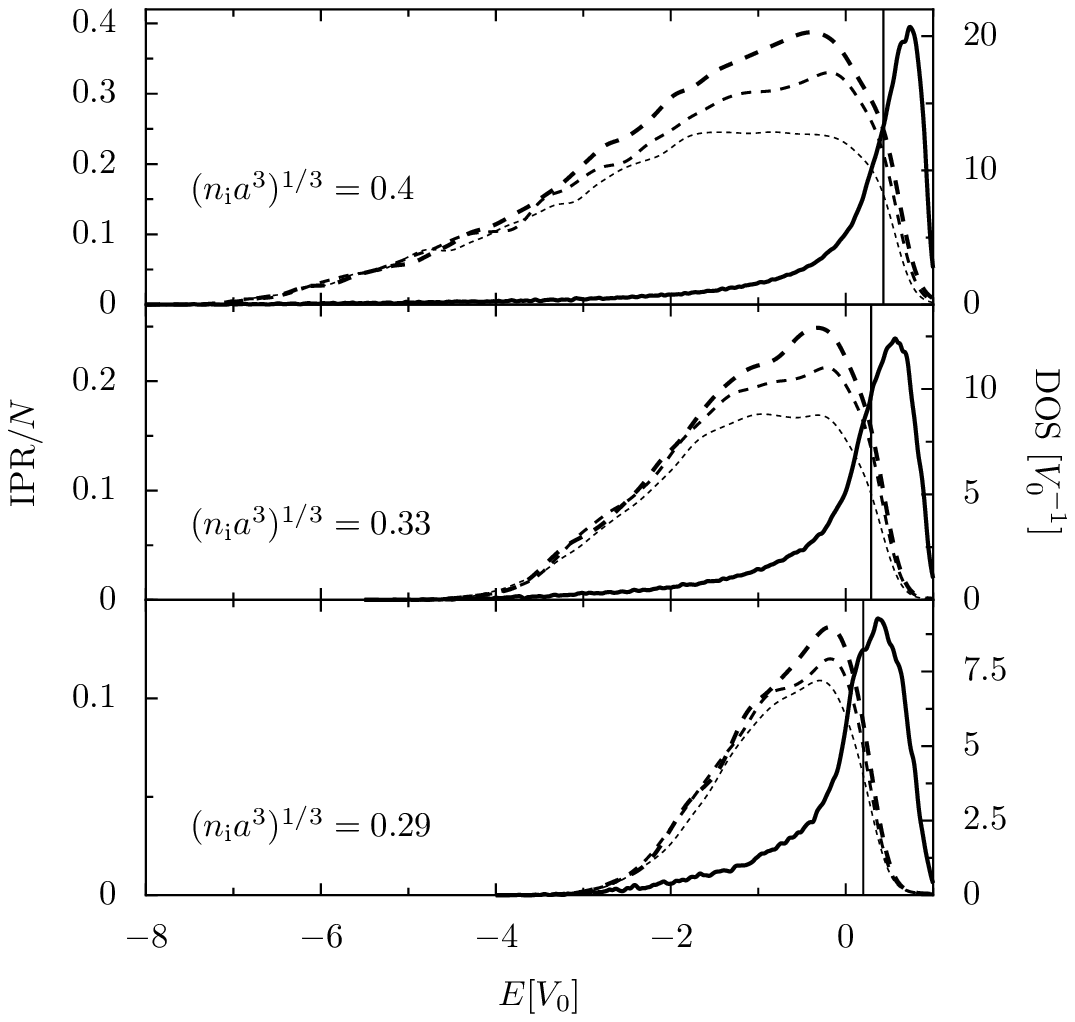}
\caption{Density of states (DOS, solid line and right scale) and 
inverse participation ratio (IPR, dashed lines and left scale) for three 
different densities on the metallic side of the metal-insulator transition. 
Dashed lines with increasing thickness are for $R_r=50$, $150$ and $250$, 
respectively. The vertical lines indicate the Fermi energy.}
\label{fig:IPRDOSMTSO}
\end{figure}

In Fig.\ \ref{fig:IPRDOSMTSO} we present the DOS and $\mbox{IPR}/N$ 
of the extended model for the three densities previously treated and 
various values of the spin-orbit coupling strength $R_r$.
The DOS does not noticeably change with $R_r$, and that is why we
only present the $R_r=1$ case.
The spin-orbit coupling results in the increase of the $\mbox{IPR}/N$ as
a function of $R_r$ in the region of extended states.
This effect is more prominent for the larger density. The low-energy sector
that has localized states in the MT model exhibits $\mbox{IPR}/N$ curves
approximately independent of $N$, which is a signature of the delocalization
tendency.

\begin{figure}
\includegraphics[width=\linewidth]{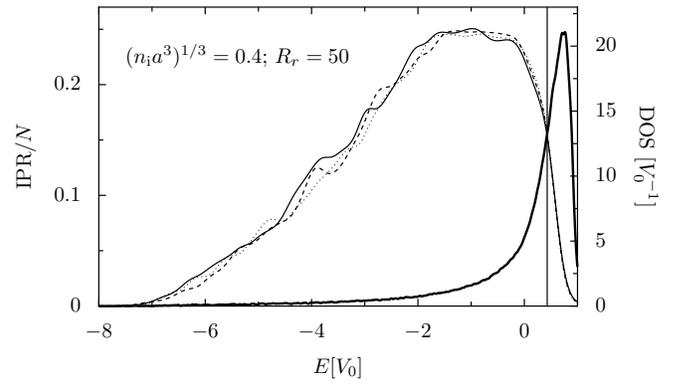}
\caption{Density of states (DOS, thick line and right scale) and inverse 
participation ratio (IPR, left scale) 
for a density on the metallic side of the metal-insulator transition, 
three different system sizes, and a fixed spin-orbit enhancement factor of 
$R_r=50$. The solid, dashed and dotted curves of $\mbox{IPR}/N$ are for 
$N=2744, 3375$ and $4096$, respectively, and the vertical line indicates the 
Fermi energy.}
\label{fig:IPRDOSMTSOsa}
\end{figure}

In Fig.\ \ref{fig:IPRDOSMTSOsa} we perform a finite-size scaling of the 
$\mbox{IPR}/N$ for a given density above the MIT critical density 
and one value of the spin-orbit coupling enhancement factor, $R_r=50$.

In conclusion, we revisited the problem of the characterization
of the eigenstates of the Matsubara-Toyozawa model from the point
of view of their localization, and performed a similar analysis
in an extended model proposed recently which includes the structural
inversion asymmetry spin-orbit mechanism. 
Analyzing the effect of spin-orbit coupling of various strengths is 
necessary in order to address the study of the spin dynamics in the 
impurity band of doped semiconductors.
We found that while the density of states is not considerably modified 
by the spin-orbit interaction, the nature of the states is noticeably 
affected by it showing a tendency to the delocalization.

We acknowledge the financial support of the Coll\`ege Doctoral
Europ\'een of the Universit\'e de Strasbourg and of UBACYT through 
grant number X495.

\end{document}